\documentclass[aps,prd,twocolumn,preprintnumbers,superscriptaddress]{revtex4-1}

\usepackage[colorlinks=true,citecolor=blue,linkcolor=blue]{hyperref}
\usepackage[normalem]{ulem}
\usepackage{amsmath,amssymb}
\usepackage{mathtools}
\usepackage{verbatim}
\usepackage{titlesec}               
\usepackage{epsfig}
\usepackage{graphicx}               
\usepackage{url}
\usepackage{color}
\usepackage{multirow}
\usepackage{floatrow}
\usepackage{placeins}
\usepackage[dvipsnames]{xcolor}
\usepackage{epstopdf}
\usepackage{siunitx}
\usepackage{fontawesome}
\usepackage{tikz}
\usepackage{tikz-feynman}
\usepackage{enumitem}
\usepackage[capitalize]{cleveref}
\usepackage{lipsum}
\usepackage{gensymb}
\usepackage{booktabs}               
\usepackage{xspace}                 
\usepackage{pifont}
\usepackage{marvosym }
\usetikzlibrary{shapes,arrows}
\usetikzlibrary{decorations.pathmorphing,decorations.markings}
\usetikzlibrary{snakes}
\usepackage{orcidlink}  
\usepackage{footnote}
\usepackage{xcolor}
\usepackage{fontawesome}
\newcommand{\gitlink}{\href{https://github.com/samiur06/running_PMNS_IceCube}{\textsc{g}it\textsc{h}ub~{\large\color{black}\faGithub}}\xspace}
\allowdisplaybreaks

\usepackage{bbm}
\usepackage{slashed}

\makeatletter

\renewcommand{\p@subsection}{}
\makeatother

\titleformat*{\section}{\centering\bfseries\uppercase}
\titlelabel{\thetitle\quad}
\titleformat*{\paragraph}{\bfseries}
\titlespacing*{\paragraph}{0pt}{3.25ex plus 1ex minus .2ex}{1em}

\makeatletter
\def\l@subsubsection#1#2{}
\makeatother

\newsavebox{\twosubbox}

\begin{document}

\title{Double Bangs at IceCube as a Window to the Neutrino Mass Origin}

\author{Samiur R. Mir \orcidlink{0000-0002-6531-2174} }
\email{samiur.mir@okstate.edu}
\affiliation{Department of Physics, Oklahoma State University, Stillwater, OK, 74078, USA}
\author{Carlos A. Argüelles \orcidlink{0000-0003-4186-4182}}
\email{carguelles@fas.harvard.edu}
\affiliation{Department of Physics \& Laboratory for Particle Physics and Cosmology,
Harvard University, Cambridge, MA 02138, USA}
\author{K.S. Babu \orcidlink{0000-0001-6147-5155}}
\email{kaladi.babu@okstate.edu}
\affiliation{Department of Physics, Oklahoma State University, Stillwater, OK, 74078, USA}
\author{Vedran Brdar \orcidlink{0000-0001-7027-5104}}
\email{vedran.brdar@okstate.edu}
\affiliation{Department of Physics, Oklahoma State University, Stillwater, OK, 74078, USA}


\begin{abstract}
Neutrino oscillation parameters are subject to renormalization group (RG) evolution, just like all couplings and masses of Standard Model (SM) particles. Within the SM extended with three massive neutrinos, it is well known that RG running effects in the neutrino sector are small. However, the RG running of the elements of the leptonic mixing (PMNS) matrix below the electroweak symmetry breaking scale can be enhanced in the presence of light neutrinophilic new particles. In this work, using a particular low-scale neutrino mass model as an example, and by taking into account both atmospheric and astrophysical neutrino fluxes, we show that RG running of the PMNS matrix can lead to an increased number of high-energy tau neutrino events at IceCube. This excess manifests as an increased number of spatially displaced showers called ``double bangs". We find that the number of double bangs induced by new physics through RG effects can be comparable to that arising from SM interactions of astrophysical tau neutrinos.
\end{abstract}

\maketitle

\noindent
\textbf{Introduction.}
Since the discovery of neutrino oscillations by Super-Kamiokande \cite{Super-Kamiokande:1998kpq} and SNO \cite{SNO:2002tuh}, many 
neutrino experiments have confirmed this phenomenon which implies that at least two neutrino species are massive. In the same period, a plethora of neutrino mass models have been proposed (see reviews in Refs.~\cite{King:2003jb,Cai:2017jrq}), together with respective search strategies including, e.g., colliders and lepton flavor violation experiments.
Recently, two of the authors of this work and collaborators demonstrated that a certain class of neutrino mass models can be tested at neutrino oscillation experiments, taking advantage of the RG running of the leptonic mixing (PMNS) matrix elements \cite{Babu:2021cxe} (see also \cite{Bustamante:2010bf,Babu:2022non,Ge:2023azz,Ge:2024ibn}). 

In a nutshell, energy scales associated with neutrino production and detection can be different. The former is typically the mass of a meson decaying to a neutrino, and the latter depends on the neutrino energy and the mass of the target particle.
In the expression for the neutrino oscillation probability, the PMNS matrix appears four times, twice each in association with production and detection scales. If a particular neutrino mass model features significant renormalization group (RG) running, the neutrino mass matrix will change across the scales and, in turn, there will be a mismatch between PMNS matrices at production and detection energy scales. Generally, in the Standard Model (SM) with three massive neutrinos, RG running effects of PMNS matrix elements are not significant within accessible energy ranges for production and detection~\cite{Babu:1993qv,Chankowski:1993tx,Antusch:2001ck,Casas:1999tg,Balaji:2000au,Antusch:2003kp,Antusch:2005gp,Goswami:2009yy}. To increase the magnitude of RG effects at the energy scales relevant for neutrino experiments, which is typically 
below the electroweak symmetry breaking (EWSB) scale, some of the new particles in a given model should have masses that lie between neutrino production and detection energy scales. 

Phenomenologically, in a simplified two-flavor scenario, the modified expression for the neutrino oscillation probability reads \cite{Babu:2021cxe}
\begin{align}
P_{\alpha\beta}= \sin^2(\theta_p-\theta_d)+\sin 2\theta_p \sin 2\theta_d \sin^2 \left(\frac{\Delta m^2 L}{4E_\nu} + \eta  \right)\,,
\label{eq:1}
\end{align}
where $\alpha,\beta$ are neutrino flavor indices, and $\Delta m^2$, $L$, and $E_\nu$ are the neutrino mass-squared difference, baseline, and energy, respectively. Further, $\theta_p$ and $\theta_d$ are the values of the mixing angle at production and detection scales, while $\eta$ is a CP-violating parameter which, unlike in the standard scenario, is already present in the toy model featuring two neutrino species \cite{Babu:2021cxe}. One can infer from \cref{eq:1} that, irrespective of the value of $\eta$, there is a nonvanishing flavor transition probability at a distance of $L$=$0$. The effects from novel sources of CP violation and zero-baseline transition probabilities are both crucial for this work, where we focus on studying the manifestation of RG effects at the IceCube experiment \cite{IceCube:2016zyt}.

In this Letter, we show that RG effects can increase the high-energy tau neutrino flux at neutrino telescopes compared to the prediction without new physics. We demonstrate this using the variant of the scotogenic neutrino mass model introduced in Ref.~\cite{Babu:2021cxe}.
We consider both atmospheric and astrophysical neutrinos, where the appearance of tau neutrinos from the former arises exclusively from zero-baseline flavor transitions, characteristic of scenarios with RG running as discussed above.
An excess of tau neutrinos implies that more ``double bang'' events are expected to be recorded in the detector.
These events consist of two spatially separated showers, typically displaced by $\sim 50 \text{ m/PeV}$: the first originates from a tau neutrino charged-current interaction that produces a tau lepton, while the second arises from the subsequent decay of the tau lepton after propagating a macroscopic distance. Before turning to the details of our analysis, let us note the complementarity of our approach with Refs.~\cite{Coloma:2017ppo,Coloma:2019qqj,Atkinson:2021rnp}, where new physics-induced double bangs are generated by the production of heavy right-handed neutrinos that decay after propagating macroscopic distances. In our scenario, new physics particles are not directly produced but enter only through loop diagrams that yield the RG running of the PMNS matrix.\\

\noindent
\textbf{The Model.}
We consider a  variant of the scotogenic model~\cite{Ma:2006km} with a $U(1)$ lepton number symmetry that was introduced in Ref.~\cite{Babu:2021cxe}. This model contains a new Higgs doublet $H_1$, three right-handed neutrinos $N_R$, and a complex scalar singlet $\varphi$. $H_1$ and $N_R$ are odd under a $\mathbb{Z}_2$ symmetry, while the lepton number assignments of the newly introduced fields are $(H_1,N_R,\varphi)=(0,+1,-2)$. 
The relevant part of the Lagrangian reads
\begin{equation}
\label{eq:lagrangian}
    - \mathcal{L} \supset 
 \overline{L} Y_{\nu} \tilde{H}_{1} N_{R} 
+ \frac{1}{2} \varphi \, \overline{N_{R}^{c}} Y_{N} N_{R} 
+ \text{h.c.}\,,
\end{equation}
where $Y_{\nu}$ and $Y_N$ are general complex $3\times3$ matrices; in our numerical computations, the latter is taken to be diagonal with real elements, which can be done without any loss of generality.
The scalar $\varphi$ gets a vacuum expectation value (VEV) $\langle\varphi\rangle\equiv v_\varphi$, thereby generating the right-handed neutrino mass $M_N = Y_N v_\varphi$. While the respective symmetry breaking preserves the $\mathbb{Z}_2$ symmetry, it breaks the lepton number. Lepton number may be broken explicitly by a soft term $(\mu^2 \varphi^2$ + h.c.) in the Higgs potential, in which case there is no Majoron present in the model. The active neutrinos acquire Majorana masses via a one-loop diagram, and the neutrino mass matrix reads  (in the approximation $M_N \ll M_{H,A}$) \cite{Babu:2021cxe}

\begin{align}
M_\nu \simeq \frac{\lambda v_\varphi}{16\sqrt{2}\pi^2} Y_\nu Y_N Y_\nu^T \ln\frac{M_H^2}{M_A^2}\,,
\label{eq:nu-mass-1}
\end{align}
where $\lambda$ is a nontrivial quartic coupling involving $H_1$ and the SM Higgs doublet, and $M_{H,A}$ are the masses of the scalars that are components of $H_1$, taken to be heavier than $\mathcal{O}(1\text{ TeV})$ scale.

We assume that $N_R$ and $\varphi$ are light, with masses below the EWSB scale and are therefore dynamical degrees of freedom across the energies of interest.
The beta function for the $Y_N$ matrix at energies above the $N_R$ and $\varphi$ mass reads~\cite{Babu:2021cxe}
\begin{align}
16\pi^2\frac{dY_N}{d\ln |Q|}= 4Y_N\left[Y_N^2 + \frac{1}{2}{\rm Tr}(Y_N^2)\right]\,,
\label{eq:running}
\end{align}
where $Q$ is the energy scale.

The matrix $Y_\nu$ does not exhibit RG evolution below the EWSB scale since $H_1$ is integrated out above such energies (new charged scalars are constrained to have masses above the EWSB scale~\cite{ALEPH:2013htx,CMS:2019bfg}),  and we assume the neutral scalars from $H_1$ have similar masses as the charged scalar. Therefore, for all practical purposes, the $Y_\nu$ matrix is constant, and we parametrize it using the Casas-Ibarra prescription \cite{Casas:2001sr}.

By definition, the PMNS matrix diagonalizes the neutrino mass matrix at any specified energy scale ($M_\nu^{\rm diag}=U^T M_\nu U$) and is parameterized at that scale as
\begin{align}
U =
R_{23}(\theta_{23}) \,
R_{13}(\theta_{13}, \delta) \,
R_{12}(\theta_{12}) \,
\mathrm{diag}\!\left(1, e^{i\tilde{\alpha}}, e^{i\tilde{\beta}}\right).
\label{eq:U3}
\end{align}
Here, $\theta_{ij}$ are the mixing angles, $\delta$ is the CP-violating phase,
$\tilde{\alpha}$ and $\tilde{\beta}$ are new sources of CP violation analogous to
$\eta$, which was introduced in \cref{eq:1} in the context of the simplified two-flavor scenario, and $R_{ij}$ is the rotation matrix in the $(i,j)$ plane. 

The PMNS matrix is energy-dependent, and we are interested in its values at the momentum transfer, $Q^2$, corresponding to both production, $Q^2_p$, and detection, $Q^2_d$, namely $U(Q^2_p)$ and $U(Q^2_d)$. The difference between these two matrices arises exclusively from the running of $Y_N$, which enters explicitly in the neutrino mass matrix (see \cref{eq:nu-mass-1}).

The neutrino oscillation probability in the setup with RG evolution reads
\begin{align}
P_{\alpha\beta}(L) =  
\left|\sum_i U^*_{\alpha i}(Q^2_p) U_{\beta i}(Q^2_d)\exp[-im_i^2L/2E_\nu]\right|^2\,.
\label{eq:Pab}
\end{align}
This formula reduces to the standard one in the case where $U(Q^2_p)=U(Q^2_d)$ and to Eq.~(\ref{eq:1}) in the case of two families.
The mismatch between $U(Q^2_p)$ and $U(Q^2_d)$ can lead to observable differences in  $P_{\alpha\beta}$ compared to the standard case with no RG running effects.

In this work, we are particularly interested in cases where 
$L \gg L_{\text{osc}}=4\pi E/\Delta m ^2$ and $L=0$, which correspond to the propagation of high-energy astrophysical and atmospheric neutrinos, respectively. 
Note that, while at $\mathcal{O}(\text{GeV})$ neutrino energies, oscillation effects at IceCube can be studied with a zenith angle-dependent baseline~\cite{IceCube:2017lak,IceCubeCollaboration:2023wtb}, at $\mathcal{O}(10^5)$ GeV energies, IceCube is effectively a zero-baseline experiment for atmospheric neutrinos, given the small $L/E$.

In the $L \gg L_{\text{osc}}$ case, we have 
\begin{align}
P_{\alpha\beta}(L) =  
\sum_i\left| U_{\alpha i}(Q_p^2)\right|^2 \left| U_{\beta i}(Q_d^2)\right|^2\,,
\label{eq:long}
\end{align}
while in the zero-baseline limit, \cref{eq:Pab} reduces to
\begin{align}
P_{\alpha\beta}(L=0) =  
\left|\sum_i U^*_{\alpha i}(Q^2_p) U_{\beta i}(Q^2_d)\right|^2 \,.
\label{eq:zero}
\end{align}
We observe from~\cref{eq:zero} that the mismatch between $U(Q^2_p)$ and $U(Q^2_d)$ can lead to flavor transitions, in contrast to the standard case where $P_{\alpha\beta}(L=0)=\delta_{\alpha\beta}$. \\

\begin{figure*}[t!]
  \centering
  \begin{minipage}{0.49\textwidth}
    \centering
    \includegraphics[width=\textwidth]{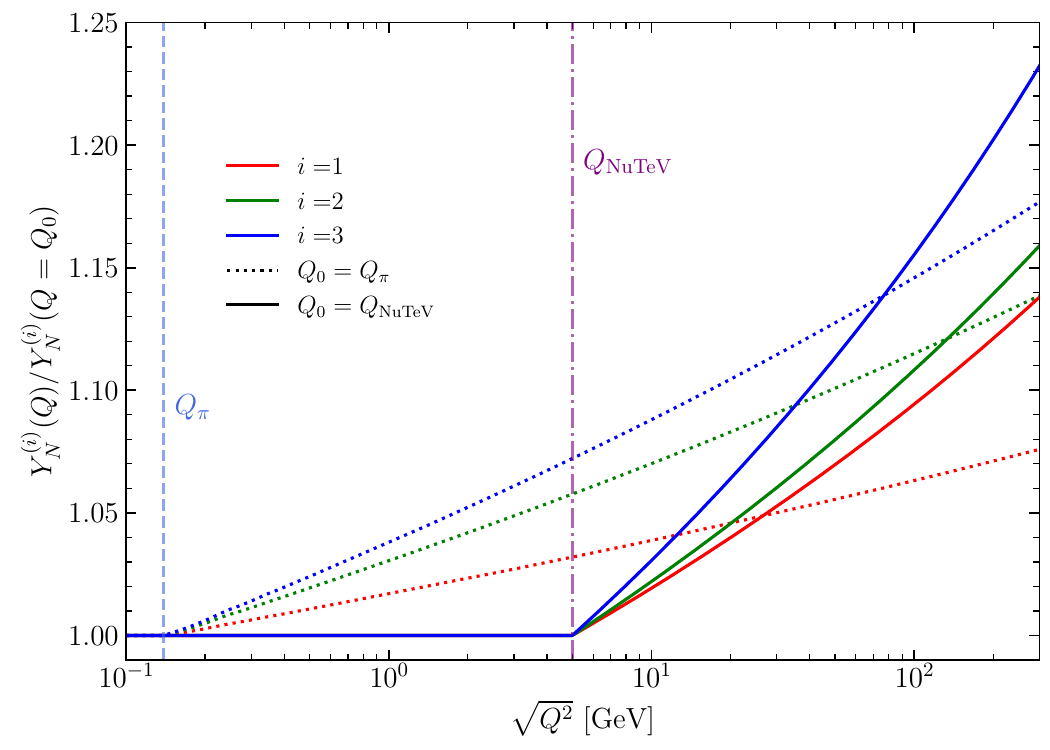}
  \end{minipage}%
  \hfill
  \begin{minipage}{0.49\textwidth}
    \centering
    \includegraphics[width=\textwidth]{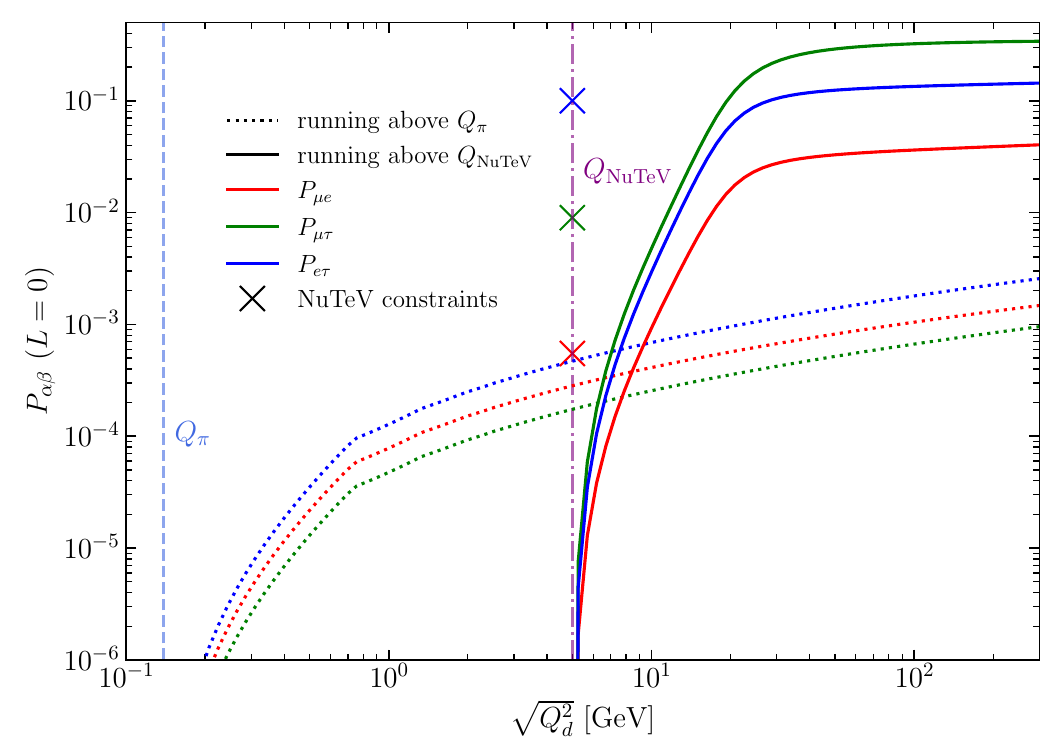}
  \end{minipage}
  \caption{The RG evolution of the $Y_N$ (left panel) and the corresponding short-baseline neutrino transition probabilities (right panel) are shown for two benchmark points in the model (solid and dotted). The $\times$ symbols in the right panel represent the NuTeV constraints on $P_{\mu e}$, $P_{\mu \tau}$, and $P_{e \tau}$~\cite{Naples:1998va,Avvakumov:2002jj}, and the vertical dashed and dot-dashed lines indicate the pion mass scale, and the approximate momentum transfer for neutrino scattering in NuTeV, respectively.}

  \label{fig:run_YN_P_abovebelowNuTeV}
\end{figure*}

\noindent
\textbf{RG running.}
In what follows, we discuss the energy range over which the RG running in the model is performed. The running of $Y_N$ starts above the mass of $N_R$ and $\varphi$; this matrix will not run below their mass scale, where these fields are integrated out. In Ref.~\cite{Babu:2021cxe}, $N_R$ and $\varphi$ were assumed to have masses equal to the pion mass $m_\pi$, which is the scale typically associated with neutrino production. Such light right-handed neutrinos are not constrained in the considered model because of the unbroken $\mathbb{Z}_2$ symmetry that prevents active-sterile neutrino mixing. 

In the left panel of \cref{fig:run_YN_P_abovebelowNuTeV},
we show with dotted lines the RG evolution for the diagonal matrix elements of $Y_N$  ($i$ labeling the diagonal element)
for a benchmark point where the RG running starts at the pion mass scale. In the right panel, for the same benchmark point and also shown with dotted lines, we present the neutrino transition probabilities $P_{\mu e}$, $P_{\mu \tau}$, and $P_{e \tau}$ in the zero-baseline limit. These probabilities, computed using \cref{eq:zero} with $|Q_p| = m_\pi$, can be used to constrain the model \cite{Babu:2021cxe} through comparison with data from short-baseline oscillation experiments such as NOMAD~\cite{NOMAD:1997pcg} and NuTeV~\cite{NuTeV:1998wnx}.
Specifically, the NuTeV experiment, which used a 250~GeV neutrino beam, reported the strongest constraints on zero-baseline transition probabilities~\cite{Naples:1998va,Avvakumov:2002jj} (see also Table~I in Ref.~\cite{Babu:2021cxe}). These constraints are indicated with $\times$ symbols in the right panel of \cref{fig:run_YN_P_abovebelowNuTeV} and placed at $Q_\text{NuTeV} \sim 5$~GeV, which corresponds to the approximate momentum transfer for neutrino scattering in this experiment. For the same benchmark point, we observe from the right panel that, for IceCube operating at larger values of $Q_d$, the transition probabilities reach at most $P_{\alpha\beta} \sim \mathcal{O}(10^{-3})$. There exist benchmark points with $|Q_p| = m_\pi$ that can produce larger transition probabilities, but these are excluded by NuTeV.

Now, if we flip the script and consider $N_R$ and $\varphi$ with masses equal to $Q_\text{NuTeV}$, $Y_N$ would be subject to RG running only above $Q_\text{NuTeV}$. In such a setup, where we effectively have $U(Q_\text{NuTeV}^2)=U(Q_p^2)$ and the mismatch takes place when $Q>Q_\text{NuTeV}$, NuTeV and other short-baseline experiments do not impose any constraints since nonvanishing zero-baseline oscillation probabilities are only generated above their typical momentum transfer scales for neutrino scattering. 
At such values of $Q_d$, we will focus on high-energy atmospheric neutrinos and astrophysical neutrinos at neutrino telescopes. 
The zero-baseline discussion applies to the former. 
For definiteness, we will focus on IceCube, since its flavor efficiencies are publicly available for high-energy events.
A relevant benchmark point for the IceCube High-Energy Starting Events (HESE)~\cite{IceCube:2020wum} selection is shown in~\cref{fig:run_YN_P_abovebelowNuTeV} with solid lines.
All the benchmarks are tabulated in the appendix for the purpose of reproducibility of these results. We observe that $P_{\alpha\beta}$ (solid) can reach $\mathcal{O}(0.1)$ without violating the NuTeV constraints, and such values are large enough to record an observable impact of the running, as we will demonstrate in the remainder of this work.

\noindent
\textbf{Neutrino Fluxes and Detector Signatures.}
In this analysis, we will employ both atmospheric and astrophysical neutrinos with energies above $10$ TeV detected using the HESE selection~\cite{IceCube:2020wum}.
Moreover, neutrinos and antineutrinos are treated identically since, modulo Glashow resonance~\cite{IceCube:2021rpz}, they cannot be distinguished at IceCube.
For the atmospheric neutrino flux, the model from Ref.~\cite{Honda:2006qj} is adopted. 
Atmospheric neutrinos with $E_\nu= 10$~TeV have an oscillation length of $L_{\text{osc}}\sim10^7$ km, which exceeds Earth's diameter by several orders of magnitude, making IceCube essentially a short-baseline experiment for high-energy atmospheric neutrinos.
Therefore, any flavor transition $\nu_\alpha \to \nu_\beta$ at such energies is effectively flavor transition at $L=0$. 
While in the standard case $P_{\alpha\beta}(L=0)=\delta_{\alpha\beta}$, the case with RG running features zero-baseline flavor transitions, see  \cref{eq:zero}. 

IceCube has by now observed hundreds of high-energy neutrinos arriving from extragalactic  sources~\cite{IceCube:2013low,IceCube:2015qii,IceCube:2020wum}.
Their flux is typically modeled as a single power law with the spectral index in between $2$ and $3$~\cite{IceCube:2020wum}. 
Irrespective of the flavor composition at the sources, the flavor ratio at Earth is approximately $\nu_e:\nu_\mu:\nu_\tau$ = $1:1:1$ in the standard scenario, given the measured values of the neutrino mixing angles~\cite{Esteban:2024eli}. In the case with RG running, \cref{eq:long} should be employed for propagating neutrinos from their production sites to Earth, and any significant excess or deficit of a measured neutrino flavor relative to the case with no RG running can be used to constrain the parameters of the model.

Regarding high-energy neutrino detection at IceCube, all three neutrino flavors can be distinguished: the signatures in the detector left by $\nu_{e}$, $\nu_{\mu}$, and $\nu_{\tau}$ map into three morphological categories: single cascades, tracks, and double bangs.
Cascades and tracks are a combination of the three flavors; we will thus focus on double bangs, which are specific to the tau flavor.
Double bangs are spatially separated showers arising from tau lepton production and decay. 
Given the magnitude of RG effects, tau neutrinos are by far the most promising for this analysis, since they do not exhibit large atmospheric neutrino backgrounds unlike $\nu_{e}$ and $\nu_{\mu}$, and thus even an $\mathcal{O}(1)$ modification of the predicted number of events for tau neutrinos can be significant.
In our analysis, we will therefore focus on $(i)$ the atmospheric $\nu_{e,\mu} \to \nu_\tau$ and $(ii)$ the astrophysical $\nu_\alpha \to \nu_\tau$ transitions.
In HESE, IceCube reported $4$ tau neutrino-candidates from $102$ detected high-energy neutrinos~\cite{IceCube:2020wum}.
Of these $4$, two were above $60~\text{TeV}$, where the atmospheric background is significantly lower.
Of the two, one was identified to have a high likelihood of being a tau neutrino~\cite{IceCube:2020fpi}.

\noindent
\textbf{Analysis and results.}
In what follows, we employ the publicly available HESE Monte Carlo (MC) sample in the IceCube HESE Data Release~\cite{IceCube:2020wum}. We first outline the parameter scan.
In the analysis, we perform a scan in the model's parameter space. The free parameters, set at $Q_p^2$ scale, are the diagonal elements of the $Y_N$ matrix taken to be of $\mathcal{O}(1)$, the 3 angles representing rotations which generate the orthogonal matrices in the Casas-Ibarra parametrization, the $3$ mixing angles and the $2$ mass squared differences that we take according to Ref.~\cite{Esteban:2024eli} and CP phases $\delta$, $\tilde \alpha$, and $\tilde \beta$, which we allow to attain any value in their physical range. For each generated parameter point, the respective neutrino transition probability can be computed, either in the zero-baseline limit (\cref{eq:zero}) for atmospheric neutrinos or by using \cref{eq:long} for astrophysical ones. The parameters are chosen such that the global fit to neutrino oscillation data is realized assuming normal ordering of neutrino masses. It should be noted that the RG effects are ineffective for reactor, solar, accelerator, and low-energy atmospheric neutrinos with our choice of $U(Q_p^2) = U(Q^2_{\rm NuTeV})$.

\begin{figure}[h!]
  \centering
    \centering
    \includegraphics[width=\textwidth]{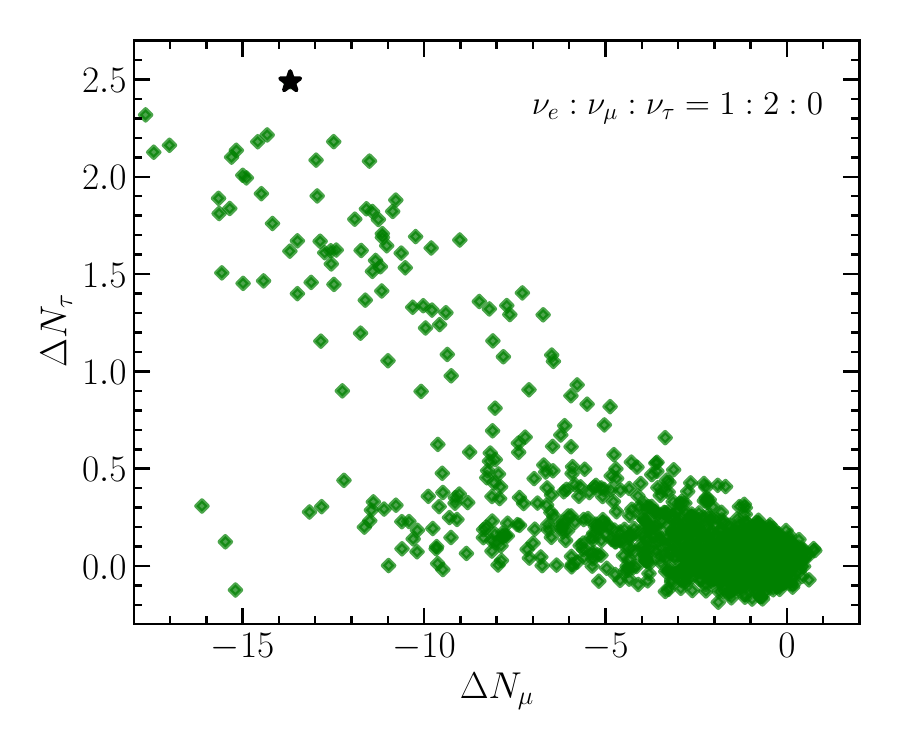}
  \caption{Difference in muon and tau neutrino event counts with and without RG running,  based on 7.5 years of IceCube data for $\mathcal{O}(10^3)$ points in the parameter space of the model. Among these, the star ($\bigstar$) marks the benchmark point that produces the maximum enhancement of predicted double bang events, $\Delta N_\tau = 2.5$, in the presence of RG running.}
  \label{fig:count_diff120}
\end{figure}

The number of tau neutrino events reads
\begin{equation}
    N_{\tau} = T \int dE_\nu \,\,\epsilon_{\tau}(E_\nu)\,\Phi_{\alpha} (E_\nu) \, \sigma_{\nu N} (E_\nu) \, w_{\alpha \to \tau}^{\mathrm{BSM}}(E_\nu)\,,
    \label{eq:count_integration}
\end{equation}
where $T$ is the data-taking time, $\epsilon_{\tau}(E_\nu)$ is the detector efficiency for identifying $\nu_\tau$, $\Phi_{\alpha}$ is the neutrino flux (either atmospheric or astrophysical), and $\sigma_{\nu N}$ is the neutrino-nucleus deep inelastic scattering (DIS) cross section.
Further, $w_{\alpha \to \tau}^{\mathrm{BSM}}$ is what we denote as “BSM weight,” and it is closely related to the neutrino flavor transition probability. 
In fact, if \cref{eq:zero,eq:long} were not $Q^2$-dependent, this factor would have been exactly equal to the respective transition probabilities.
However, given the momentum transfer dependence in $P_{\alpha\tau}$, the transition probabilities need to be weighted with the $Q^2$-dependent differential cross section $\frac{d^{2}\sigma_{\nu(\bar{\nu})}}{dx \, dQ^{2}}$ ~\cite{Formaggio:2012cpf,Xie:2023suk}, and this is precisely how the BSM weight is computed. The isoscalar $u (\bar{u})$ and $d (\bar{d})$ PDFs, that appear in the differential cross section, are calculated from CT18 NNLO proton PDFs~\cite{Hou:2019efy}. For details of the BSM weight, we refer the reader to the appendix.

\begin{figure*}[t!]
  \centering
  \begin{minipage}{0.49\textwidth}
    \centering
    \includegraphics[width=\textwidth]{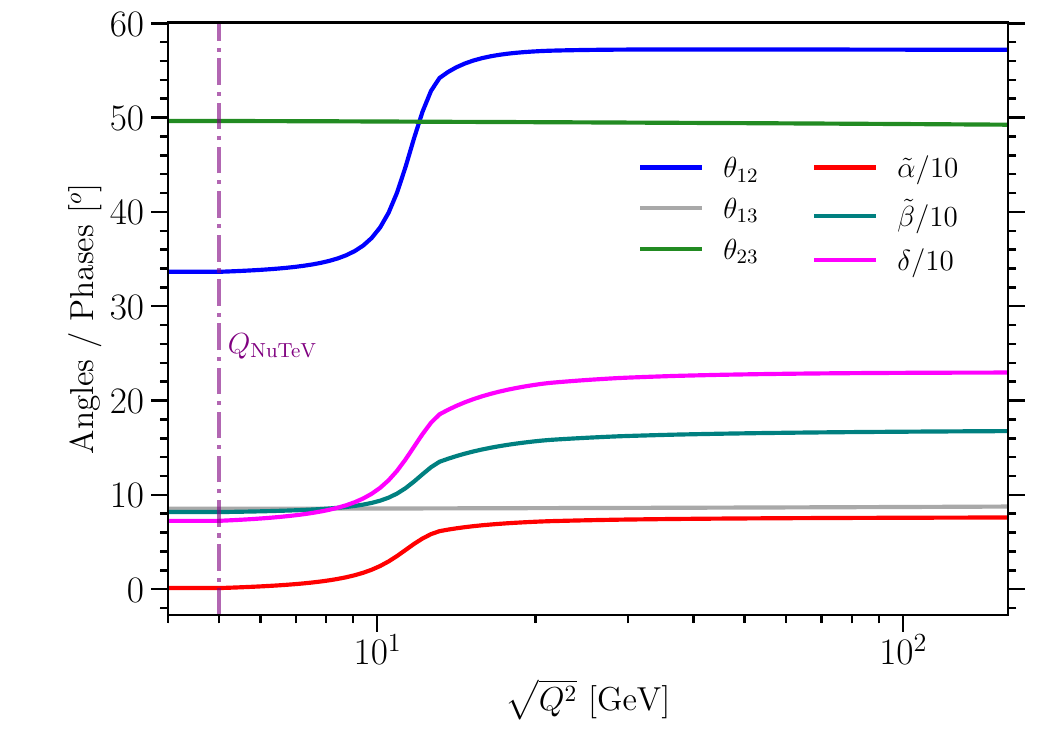}
  \end{minipage}%
  \hfill
  \begin{minipage}{0.49\textwidth}
    \centering
    \includegraphics[width=\textwidth]{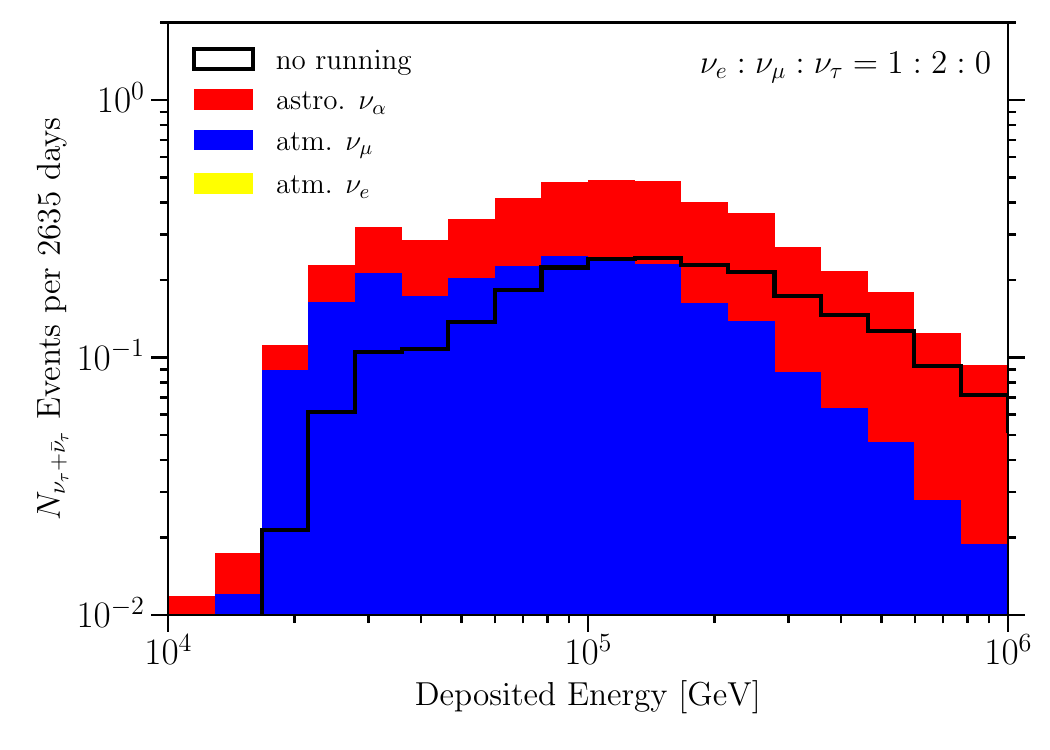}
  \end{minipage}
  \caption{In the left panel, we show the RG evolution of mixing angles and phases as defined in \cref{eq:U3}. In the right panel, 
  we show the corresponding reconstructed energy spectrum of the tau-like events.}
  \label{fig:star_benchmark}
\end{figure*}

For atmospheric neutrinos, in the language of the HESE data release associated with Ref.~\cite{IceCube:2020wum}, \cref{eq:count_integration} effectively becomes
\begin{align}
N^{\rm atm}_{\tau}
  = T \sum_{i}
      w_{i,\mathrm{MC}}^{\tau}\, \sum_{\alpha=e, \mu}
      \Phi_{\alpha} ^{\textrm{atm}} (E^{0}_{\nu, i}, \theta^{0}_i)\,
      w_{\alpha \to \tau}^{\mathrm{BSM}, i} (E_{\nu, i})\,,
\label{eq:atm_count}
\end{align}
where the integral is replaced by the sum over events in the MC sample.
Here, $\alpha$ stands for electron and muon neutrinos since tau neutrinos are not produced in the atmosphere, $E_{\nu, i}$ is the reconstructed deposited energy, and ${E^0_{\nu, i}, \theta^0_i}$ are the energy and zenith angle of the $i$-th event. We fix $T=7.5$ years according to Ref.~\cite{IceCube:2020wum}. The cross section and the detection efficiency are now part of the event weight $w_{i,\mathrm{MC}}^\tau$. The incoming flux and detector efficiency hold for the respective neutrino flavor in the MC events provided through the data release, hence we calculate the flux $\Phi_{\alpha\neq\tau}^\text{atm}$ using the \verb|`honda2006'| model from the \verb|nuflux| package~\cite{IceCube_Collaboration_NuFlux_A_library_2024}. Therefore, note that true parameters, instead of reconstructed ones, for the energy and zenith angle are used for the incoming flux.
Finally, using the data release~\cite{IceCube:2020wum}, we only select the tau neutrino-like MC events associated with double bang topology, requiring a minimum separation of $10$ m between the generated showers~\cite{IceCube:2015vkp,IceCube:2020wum}.

The treatment of tau neutrino events from the astrophysical flux centers around the flavor ratio, as it can deviate from the standard one $\nu_e:\nu_\mu:\nu_\tau =1:1:1$ due to RG running. The number of tau neutrino events from the astrophysical flux reads
\begin{equation}
\label{eq:astro_count}
{N^{\rm astro}_{\tau } = T \sum_{i} w_{i,\mathrm{MC}}^{\tau}\,\Phi^{\rm astro}_{1 \nu} (E_{\nu, i})\,
      \sum_{\alpha=e, \mu, \tau} w_{\alpha \to \tau}^{i,\mathrm{BSM}} (E_{\nu, i})  X_{\alpha}^{\rm prod},}
\end{equation}
where $X_{\alpha}^{\rm prod}$ represents the flavor fraction of a given neutrino type at the production site, and  $\Phi^{\rm astro}_{1 \nu} (E_{\nu})=\frac{1}{6}\Phi^{\rm astro}_{6 \nu} (E_{\nu})$, is the (anti)neutrino flux for a single flavor, obtained from the single power law formula from~\citep{IceCube:2020wum}.

For a given parameter point, including both atmospheric and astrophysical neutrino fluxes, we can compare the expected number of double bangs in the presence of RG running with the result in the case with no RG effects; the latter is computed by setting BSM weights $w_{\alpha \to \tau}^{\mathrm{BSM}}$ to ${\delta_{\alpha \tau}}$. We did this for $\mathcal{O}(10^3)$ parameter points in the model, and the results are shown in \cref{fig:count_diff120}, displaying the difference in number of events between the running and the standard case, i.e., $\Delta N_\alpha = N^{\rm BSM} _{\alpha} - N^{\rm std} _{\alpha}$. All parameter points, event differences, and the results for different initial flavor compositions of astrophysical neutrinos are available on~\gitlink. 
We found that the results are only mildly dependent on the flavor ratio at the production site; we present $\nu_e:\nu_\mu:\nu_\tau = 1:2:0$ in \cref{fig:count_diff120}. On the y-axis, we show an enhancement of double bangs with respect to the standard expectation, $\Delta N_\tau$. On the x-axis, for illustration, we show the difference in $\nu_\mu$-like track events for the corresponding parameter points, computed analogously to tau events (see \cref{eq:atm_count,eq:astro_count}), only with different weights to account for different transition probabilities and detection efficiencies. The $\Delta N_\mu$ values reported in \cref{fig:count_diff120} are consistent with the measurement of the atmospheric muon neutrino flux at IceCube \cite{IceCube:2014slq}.  We observe a number of parameter points for which RG running yields $\sim2$ additional double bangs, with a single parameter point reaching $\Delta N_\tau=2.5$ (labeled with a star).

For this benchmark point, we show the RG evolution for mixing angles and phases as well as the reconstructed energy spectrum of tau-like events in \cref{fig:star_benchmark}. The angle $\theta_{12}$, along with all three phases, changes significantly in the relevant $Q^2$ range in the left panel of \cref{fig:star_benchmark}, while the angles $\theta_{13}$ and $\theta_{23}$ do not, in accord with the arguments in Ref.~\cite{Antusch:2005gp}. The right panel compares the case with no running to the case where additional double bang events arise from atmospheric and astrophysical neutrinos in the presence of RG effects. The atmospheric contribution from $\nu_e$ is too small to be observed in the presented range. If one subtracts the unfilled (no running) histogram from the stacked one and integrates over energies, $\Delta N_\tau = 2.5$ from \cref{fig:count_diff120} is obtained.

By comparing $\Delta N_\tau = 2.5$ with the expected number of double cascade events in HESE MC sample~\cite{IceCube:2020wum}, which is $\sim 2.5$, we observe that RG running can increase the number of double bang events compared to the standard scenario by up to 100\%. 
This implies that RG effects are already testable with IceCube, and they will be even more prominent in the era of IceCube-Gen2~\cite{IceCube-Gen2:2020qha}, where the enhancement of double bang events due to new physics can be probed with larger sample sizes.

\noindent
\textbf{Conclusions.}
In this work, we have shown for the first time that neutrino telescopes can be used to constrain a particular class of neutrino mass models that exhibit non-negligible RG running effects at low energies. 
For high-energy atmospheric neutrinos, these effects induce zero-baseline transition probabilities.
Similarly, for astrophysical neutrinos, RG running modifies the expected neutrino flavor composition at Earth. 
Combining these effects, we have demonstrated that characteristic signatures of high-energy tau neutrinos, the so-called double bang events, provide a particularly promising avenue for constraining mass models featuring RG running. 
Focusing on a specific scotogenic-like model, we found that RG effects can increase the number of expected double bang events at neutrino telescopes by up to 100\%. 
Beyond IceCube-Gen2, our work motivates the importance of measuring the number of tau neutrinos at the highest energies, which further motivates specialized tau-neutrino telescopes such as TAMBO~\cite{TAMBO:2025jio} and Trinity~\cite{Otte:2025dld} in the sub-100~PeV range and GRAND~\cite{GRAND:2018iaj}, Beacon~\cite{BEACON:2025qcq}, and other radio detectors at even higher energies~\cite{Ackermann:2022rqc}.
In summary, our work opens a new direction for probing the origin of neutrino mass, complementing and extending existing approaches.\\


\textbf{Acknowledgements.} SRM would like to thank Soumyananda Goswami for valuable discussions. The work of VB and SRM is supported by the United States Department of Energy under Grant No. DE-SC0025477. The work of KSB is supported in part by the U.S. Department of Energy under grant number DE-SC0016013. CAA is supported by the Faculty of Arts and Sciences of Harvard University, the National Science Foundation, the Research Corporation for Science Advancement, and the David \& Lucile Packard Foundation. 
VB acknowledges support from the Oklahoma State University College of Arts and Sciences through its Fall Travel Program. VB and SRM gratefully acknowledge the Laboratory for Particle Physics and Cosmology at Harvard University for its generous hospitality, where part of this research was carried out. We acknowledge the use of the High Performance Computing Center at Oklahoma State University, which is supported in part by the National Science Foundation under Grant No. OAC-1531128.

\bibliographystyle{JHEP}
\bibliography{refs}

\clearpage
\newpage
\onecolumngrid
\appendix




\section*{Appendix}

\section*{1. BSM weights}

At the considered energies, the neutrino-nucleus interaction is governed by deep inelastic scattering (DIS). A neutrino $\nu_\alpha$ can produce $\ell_\alpha$ through the charged-current interaction in which a $W$ boson is exchanged. 
The flavor-independent charged-current DIS differential cross section reads~\cite{Formaggio:2012cpf,Xie:2023suk}
\begin{equation}
\label{eq:diff_DIS_xsec}
    \frac{d^{2}\sigma_{\nu(\bar{\nu})}}{dx \, dQ^{2}}
= \frac{G_{F}^{2}}{4 \pi x \left(1 + Q^{2}/M_{W}^{2}\right)^{2}}
\left[ Y_{+} F_{2}  - y^{2} F_{L} \pm Y_{-} x F_{3} \right],
\end{equation}
where $F_{2,3,L}$ are structure functions, $Y_{\pm}$ are functions of kinematic invariants, and $x$ and $y={Q^2}/{(2xm_N E_\nu)}$ are Bjorken variables. 
We employ the DIS cross section (up to leading order, $F_L=0$) to compute the BSM weights $w_{\alpha \to \beta}^{\mathrm{BSM}}(E_\nu)$, in order to properly treat the RG evolution–modified flavor transitions. 
Specifically, we define the BSM weight as the ratio of the integrated DIS differential cross section weighted by the neutrino transition probability to the standard DIS cross section
\begin{equation}
\label{eq:wBSM}
w_{\alpha \to \beta}^{\mathrm{BSM}}(E_\nu)
  = \frac{\int_{\nu} \,   P_{\alpha \beta}(E_\nu,Q^2)      + \int_{\bar{\nu}}\,  P_{\bar{\alpha} \bar{\beta}}(E_\nu, Q^2)}{\int_{\nu} \quad
  + \quad \int_{\bar{\nu}}   }\,.
\end{equation}
The integration $\int_{\nu(\bar{\nu})}$ evaluated with $Q_{\text{max}}^2=2m_N E_\nu$ and $Q_{\text{min}}=1.3~\text{GeV}$ (CT18 NNLO) follows
\begin{equation}
    \int_{\nu(\bar{\nu})}= \int_{Q_{\text{min}}^2}^{Q_{\text{max}}^2}\, dQ^2\,\int_{Q^2/Q_{\text{max}}^2} ^1\,dx\, \frac{d^2\sigma_{\nu(\bar{\nu})} } {dx\,dQ^2}\,.
\end{equation}

\section*{2. Benchmarks}
The CP-odd phases ($\tilde{\alpha}, \tilde{\beta}$, and $\delta$) are defined in \cref{eq:U3}. Their values at the production scale are free parameters. The  Casas-Ibarra parametrization yields
\begin{align}\label{eq:casas-ibarra}
Y_\nu=\frac{1}{\sqrt{C}}\,U^\dagger(Q_p^2)\,  \sqrt{\text{diag}\left(m_{\nu_1},m_{\nu_2},m_{\nu_3}\right)}\,R\, Y_N^{-1/2}\,,
\end{align}
where $C$ is the VEV-dependent prefactor in~\cref{eq:nu-mass-1}. The orthogonal matrix $R$ is computed by three Euler-like rotations, $R=R_{12}(\xi_1) R_{23}(\xi_2) R_{13}(\xi_3)$ with rotation angles $\xi_1$ in the 1-2 plane, $\xi_2$ in the 2-3 plane, and $\xi_3$ in the 1-3 plane. In addition to $\xi_i$ and $Y_N$, the lightest neutrino mass enters as a free parameter in the diagonal mass matrix in \cref{eq:casas-ibarra}.  
We present the model parameters (rounded to three significant digits) at the production scale. They correspond to three benchmark points. Two are used in \cref{fig:run_YN_P_abovebelowNuTeV} (dotted and solid lines) and are labeled here as `$Q_\pi$'  and `$Q_{\text{NuTeV}}$', respectively. The third one is the $\bigstar$ benchmark from \cref{fig:count_diff120}.

\begin{table}[h!]
\centering
\renewcommand{\arraystretch}{1.2}
\setlength{\tabcolsep}{5pt}
\begin{tabular}{|c|c|c|c|c|c|c|c|c|}
\hline
benchmark & $\sin^2\theta_{12}$ & $\sin^2\theta_{13}$ & $\sin^2\theta_{23}$ & $\tilde{\alpha}$ & $\tilde{\beta}$ & $\delta$ & $\Delta m_{21}^2~(\times10^{-5}\text{ eV}^2)$ & $\Delta m_{31}^2~(\text{eV}^2)$ \\
\hline
\hline
$Q_\pi$ & 0.307 & 0.0219 & 0.450 & 0.0766 & 0.00805 & 2.85 & $7.49$ & 0.00247 \\
\hline
$Q_{\text{NuTeV}}$ & 0.307 & 0.0219 & 0.490 & 0.101 & 1.53 & 0.599 & $7.49$ & 0.00248 \\
\hline
$\bigstar$ & 0.307 & 0.0219 & 0.581 & 0.0204 & 1.43 & 1.26 & $7.49$ & 0.00253 \\
\hline
\end{tabular}
\caption{Leptonic mixing parameters and neutrino mass-squared differences for the selected benchmark points.}
\end{table}

\begin{table}[h!]
\centering
\renewcommand{\arraystretch}{1.2}
\setlength{\tabcolsep}{5pt}
\begin{tabular}{|c|c|c|c|c|c|c|c|}
\hline
benchmark & $\xi_1$ & $\xi_2$ & $\xi_3$ & $Y_N^{(1)}$ & $Y_N^{(2)}$ & $Y_N^{(3)}$ & $m_1~(\text{eV})$ \\
\hline
\hline
$Q_\pi$ & 2.35 & 1.04 & 2.28 & 0.00886 & 0.507 & 0.627 & 0.0144 \\
\hline
$Q_{\text{NuTeV}}$ & 5.05 & 0.0984 & 4.10 & 0.523 & 0.649 & 0.938 & 0.0476 \\
\hline
$\bigstar$  & 1.71 & 0.121 & 4.66 & 0.0395 & 0.139 & 0.921 & 0.0453 \\
\hline
\end{tabular}
\caption{Rotation angles in the Casas-Ibarra parametrization, the Yukawa couplings, and lightest neutrino mass for the selected benchmark points. We consider the normal mass ordering.}
\end{table}

\end{document}